\documentclass{emulateapj}

\shorttitle{Simulations of Winds of Weak-Lined T Tauri Stars}
\shortauthors{Vidotto et al.}

\begin{document}

\title{Simulations of Winds of Weak-Lined T Tauri Stars: The Magnetic Field Geometry and The Influence of the Wind on Giant Planet Migration}

\author{A. A. Vidotto}
\affil{University of S\~ao Paulo, Rua do Mat\~ao 1226, S\~ao Paulo, SP, Brazil, 05508-090} 
\affil{George Mason University, 4400 University Drive, Fairfax, VA, USA, 22030-4444}
\email{aline@astro.iag.usp.br}              

\author{M. Opher}
\affil{George Mason University, 4400 University Drive, Fairfax, VA, USA, 22030-4444}

\author{V. Jatenco-Pereira}
\affil{University of S\~ao Paulo, Rua do Mat\~ao 1226, S\~ao Paulo, SP, Brazil, 05508-090}

\and

\author{T. I. Gombosi}
\affil{University of Michigan, 1517 Space Research Building, Ann Arbor, MI, USA, 48109-2143}

\begin{abstract}
By means of numerical simulations, we investigate magnetized stellar winds of pre-main-sequence stars. In particular we analyze under which circumstances these stars will present elongated magnetic features (e.g., helmet streamers, slingshot prominences, etc). We focus on weak-lined T Tauri stars, as the presence of the tenuous accretion disk is not expected to have strong influence on the structure of the stellar wind. We show that the plasma-$\beta$ parameter (the ratio of thermal to magnetic energy densities) is a decisive factor in defining the magnetic configuration of the stellar wind. Using initial parameters within the observed range for these stars, we show that the coronal magnetic field configuration can vary between a dipole-like configuration and a configuration with strong collimated polar lines and closed streamers at the equator (multi-component configuration for the magnetic field). We show that elongated magnetic features will only be present if the plasma-$\beta$ parameter at the coronal base is $\beta_0 \ll 1$. Using our self-consistent 3D MHD model, we estimate for these stellar winds the time-scale of planet migration due to drag forces exerted by the stellar wind on a hot-Jupiter. In contrast to the findings of Lovelace et al. (2008), who estimated such time-scales using the Weber \& Davis model, our model suggests that the stellar wind of these multi-component coronae are not expected to have significant influence on hot-Jupiters migration. Further simulations are necessary to investigate this result under more intense surface magnetic field strengths ($\sim 2$-$3$~kG) and higher coronal base densities, as well as in a tilted stellar magnetosphere.
\end{abstract}


\section{INTRODUCTION}

T Tauri stars are recently formed low-mass stars ($0.5 \lesssim M/M_\odot \lesssim 2$), with spectral type ranging from F to M, and radius of no more than $3-4~R_\odot$. They are the predecessors of solar-like stars that are still in the pre-main sequence (PMS) phase. Usually, T Tauri stars are classified in two categories: classical and weak-lined. A classical T Tauri star (CTTS) shows strong emission line activity, excess ultraviolet and infrared emissions, which are associated with accretion flows and the presence of circumstellar disks. A weak-lined T Tauri star (WTTS), on the other hand, has a weaker emission line activity and shows little or no trace of accretion, implying that the accretion disk has been dissipated. For this reason, they are also known as naked T Tauri stars.

WTTSs also show radio and X-ray flare activity that is usually explained as a scaled-up magnetic activity similar to the one observed in the Sun. In the Sun, highly detailed observations of the magnetic field, using both {\it in situ} and remote measurements, have shown a very complex magnetic configuration \citep{1995JGR...10019893M,2005ApJ...622.1251L, 2008A&A...488..303S}. Although in less detail, magnetic activity has also been observed in other stars \citep[e.g., ][]{2005MNRAS.361..837P, 2006ApJ...646L..73P, 2006ApJ...644..497R, 2007MNRAS.376.1145W, 2007ApJ...664..975J, 2008ApJ...683..466V, 2008MNRAS.390..545D}. As observational techniques are becoming more accurate, detection of a more complex magnetic field geometry in young stars has just started to be accomplished, for example, the CTTSs V2129 Oph \citep{2007MNRAS.380.1297D}, BP Tau \citep{2008MNRAS.386.1234D}, and CV Cha \& CR Cha \citep{2009arXiv0905.0914H} where surface magnetic maps have been derived from spectropolarimetric data. Another system, V773 Tau A, a young binary system consisted of two WTTSs, is known to posses high levels of magnetic activity \citep{1991ApJ...382..261P, 1996AJ....111..918P}. Recently, \citet{2008A&A...480..489M} observed two extended radio emissions that were interpreted as associated to the primary and secondary stars of the system. They suggested that these features were produced by helmet streamer structures from each of the stars, that when overlapped at certain orbital phases, would result in reconnection producing the giant flares that were observed: radio emission would increase from a few mJy at apastron to more than 100 mJy at periastron. In another recent work, \citet{2008MNRAS.385..708S} reported the findings of {slingshot prominences} on TWA6, another WTT star. In their interpretation, such magnetic prominence extends for about $4$ stellar radii, which is beyond the corotation radius located at $2.4$ stellar radii. Magnetic field extrapolations from surface magnetograms have also shown that the geometry of the large-scale field of T Tauri stars may be very complex \citep{2008MNRAS.386..688J, 2008MNRAS.389.1839G}.

In the present work, we aim to investigate the magnetic configuration of the corona of WTTSs using three-dimensional (3D) magnetohydrodynamics (MHD) numerical simulations. T Tauri stars are believed to be magnetized, and rotating several times faster than the Sun, hence it is important to consider a magneto-centrifugal wind. 

The first theoretical models to analyze the effects of both rotation and magnetic field on stellar winds are \citet[WD from now on]{1967ApJ...148..217W} and \citet{1968MNRAS.138..359M}, who investigated the coupling of a purely radial stellar magnetic field to the wind using MHD approach. Recently, theoretical and numerical efforts were made towards the study of more realistic magneto-centrifugal winds in different systems. For instance, \citet{1993MNRAS.262..936W} studied the influence of stellar rotation on the wind structure and acceleration of magnetized main-sequence solar-like stars investigating the effects of the centrifugal force. \citet{2005A&A...440..411H}, on the other hand, analyzed the impact of a purely radial field on the wind. The stellar field, although radial, was taken to have a non-uniform distribution in the stellar surface, as to account for the non-uniform surface magnetic flux distribution observed in active stars. \citet{2008ApJ...678.1109M} aiming to explain the low rotation rates observed in CTTSs, studied the role of an accretion powered stellar wind in the spinning down of a typical CTTS. 

All these works have confirmed that the structure of a magnetized stellar wind significantly depends on both the field intensity and topology. In \citet{paper1}, we performed an analysis of coronal winds in main-sequence solar-like stars. We showed that a decisive parameter in the acceleration of the magnetized wind of a solar-like star is the plasma-$\beta$ at the coronal base. Using 3D MHD self-consistent simulations, we explored how the change in the magnetic and thermal energy densities at the base of the wind modifies the magnetic configuration of the corona and the velocity profile of the wind. Rotation was not considered in that work.

We now investigate the magnetic configuration of the winds of WTTSs. In our models we vary different physical parameters of the systems, such as stellar rotation period, and magnetic field intensity and gas density at the base of the wind. In particular, we investigate the occurrence of helmet streamers or slingshot prominences, as well as their shape and extension. We perform self-consistent time-dependent 3D numerical simulations, including centrifugal, magnetic, gravitational, and thermal forces. The full set of MHD equations are solved to obtain the steady-state solution for the large-scale magnetic configuration of a TTauri star. 

The paper is organized as follows. In \S 2, we present the numerical scheme used, in \S 3 we describe our choice of parameters used in the simulations, and in \S 4, we present and discuss the results obtained. In \S 5, we analyze planetary migration due to drag forces exerted by the magnetized stellar wind on ``hot-Jupiters'', giant planets orbiting very close to the host star ($\lesssim 0.1$~AU). \S 6 is dedicated to conclusions and discussion.

\section{THE NUMERICAL MODEL} 
To perform the simulations, we make use of the Block Adaptive Tree Solar-wind Roe Upwind Scheme (BATS-R-US), a 3D MHD numerical code developed at the Center for Space Environment Modeling at University of Michigan \citep{1999JCoPh.154..284P}. BATS-R-US has a block-based computational domain, consisting of Cartesian blocks of cells that can be adaptively refined for the region of interest. It has been used to simulate the heliosphere \citep{2003ApJ...595L..57R}, the outer-heliosphere \citep{1998JGR...103.1889L, 2003ApJ...591L..61O, 2006ApJ...640L..71O, 2007Sci...316..875O}, coronal mass ejections \citep{2004JGRA..10901102M,2005ApJ...627.1019L}, the Earth's magnetosphere \citep{2006AdSpR..38..263R} and the magnetosphere of other planets \citep{2004JGRA..10911210T,2005GeoRL..3220S06H}, among others. In this work, we extend the model developed in \citet{paper1} to study the wind structure of WTTSs. 

BATS-R-US solves the ideal MHD equations, that in the conservative form are given by
\begin{equation}
\label{eq:continuity_conserve}
\frac{\partial \rho}{\partial t} + \nabla\cdot \left(\rho {\bf u}\right) = 0
\end{equation}
\begin{equation}
\label{eq:momentum_conserve}
\frac{\partial \left(\rho {\bf u}\right)}{\partial t} + \nabla\cdot\left[ \rho{\bf u\,u}+ \left(p + \frac{B^2}{8\pi}\right)I - \frac{{\bf B\,B}}{4\pi}\right] = \rho {\bf g}
\end{equation}
\begin{equation}
\label{eq:bfield_conserve}
\frac{\partial {\bf B}}{\partial t} + \nabla\cdot\left({\bf u\,B} - {\bf B\,u}\right) = 0
\end{equation}
\begin{equation}
\label{eq:energy_conserve}
\frac{\partial\varepsilon}{\partial t} +  \nabla \cdot \left[ {\bf u} \left( \varepsilon + p + \frac{B^2}{8\pi} \right) - \frac{\left({\bf u}\cdot{\bf B}\right) {\bf B}}{4\pi}\right] = \rho {\bf g}\cdot {\bf u} \, ,
\end{equation}
where $I$ is the identity matrix, $\rho$ is the mass density, ${\bf u}$ the plasma velocity, ${\bf B}$ the magnetic field, $p$ the gas pressure, ${\bf g}$ the gravitational acceleration due to the central body, and  $\varepsilon$ is the total energy density given by 
\begin{equation}\label{eq:energy_density}
\varepsilon=\frac{\rho u^2}{2}+\frac{p}{\gamma-1}+\frac{B^2}{8\pi} \, .
\end{equation}
We consider ideal gas, so $p=\rho k_B T/(\mu m_p)$, where $k_B$ is the Boltzmann constant, $T$ is the temperature, $\mu m_p$ is the mean mass of the particle, and $\gamma$ is the ratio of the specific heats. Due to the lack of knowledge of all the detailed processes that take place in a stellar wind, it is difficult to estimate all the mechanisms that modify the heat content of the wind (e. g., conduction, radiation, mechanical dissipation of energy that is transferred to the plasma). We adopt, therefore, an approach similar to \citet{2003ApJ...595L..57R} who considered that $\gamma$ is associated with ``turbulent'' internal degrees of freedom, in a way analogous to the Sun, where a significant amount of energy is stored in the form of waves and turbulent fluctuations.

For all the simulations performed, we adopted the same grid resolution. Initially, the simulation domain is refined in five levels. Other five refinement levels that are body-focused and focused on the equatorial plane (current sheet region) are applied next. Finally, an additional level is applied to the body. There are $9.1 \times 10^6$ cells in the domain. The smallest cell size is $0.018~R_*$, located around the central star, where $R_*$ is the stellar radius. The maximum cell size is $4.68~R_*$. The cell size near the current sheet is $0.036~R_*$. The grid is Cartesian and the star is placed at the origin. The axes $x$, $y$ and $z$ extend from $-75~R_*$ to $75~R_*$. More details of the grid can be found in \citet{paper1}.

The inner boundary of the system is the base of the wind at $r=R_*$. Fixed boundary conditions were adopted at $r=R_*$. The outer boundary has outflow conditions, i. e., a zero gradient is set to all the primary variables.

The star is located at the center of the grid and has $M_\star = 0.8~M_\odot$ and $R_*=2~R_\odot$. The grid is initialized with a 1D hydrodynamical (HD), thermal pressure driven wind for a fully ionized hydrogen plasma. This solution is dependent on the choice of the temperature at the base of the wind and on $\gamma$, and the only physical possible solution is the one that becomes supersonic when passing through the critical radius \citep{1958ApJ...128..664P}. The initial temperature profile obeys the polytropic relation $T \propto \rho^{\gamma -1}$. Due to conservation of mass of a steady wind, we obtain the density profile from the radial velocity profile $u_r (r)$. 

The star is considered to be rotating as a solid body with a period of rotation $P$. The rotation axis is along the $z$ direction, parallel to the magnetic moment vector. The simulations are initialized with a dipolar magnetic field configuration described in spherical coordinates $\{ r, \theta, \varphi \}$ by
\begin{equation}
\label{eq:dipole}
{\bf B} =  \frac{B_0 R_* ^3}{r ^3} \left(\cos \theta , \frac12 \sin \theta, 0 \right) \, ,
\end{equation}
where $B_0$ is the magnetic field intensity at the poles, $r$ is the radial coordinate, $\theta$ is the colatitude, and $\varphi$ is the azimuthal angle measured in the equatorial plane. As the magnetic energy density is latitude dependent and the thermal pressure is held constant at the base of the wind, the plasma-$\beta$ at the surface of the star is minimum at the pole. The MHD solution is evolved in time from the initial dipolar configuration for the magnetic field to a fully self-consistent non-dipolar solution, until steady state is achieved.

\section{THE CHOICE OF PARAMETERS}
In this section, we describe the choice of the parameters used in the simulations. They can be seen in Table \ref{table:4}, where we present: the maximum surface magnetic field intensity $B_0$, the stellar period of rotation $P$, the ratio of specific heats $\gamma$, the density at the base of the corona $\rho_0$, and $\beta_0$, the surface value of the plasma-$\beta$ evaluated at the pole. We chose to simulate extreme values of the stellar rotation period, magnetic field intensity, and density, in order to probe the possible magnetic configurations for WTTSs. A brief explanation of our choices is presented next.

\begin{deluxetable}{c c c c c c}   
\tablewidth{0pt}
\tablecaption{The set of simulations. The columns represent, respectively: the name of the simulation, the surface magnetic field intensity at the pole, the period of rotation of the star, $\gamma$, the density at $R_*$, and the plasma-$\beta$ evaluated at the pole. \label{table:4}}  
\tablehead{\colhead{Name} &  \colhead{$B_0$~(kG)}  & \colhead{$P$ (d)} & \colhead{$\gamma$} & \colhead{$\rho_0$} (g~cm$^{-3}$) & \colhead{$\beta_0$}}
\startdata
A &  $1$ &  $3$ & $1.2$  & $1\times 10^{-11}$ & $1/25$ \\
B &  $0.2$  &  $3$ & $1.2$  & $1\times 10^{-11}$ & $1$ \\
C &  $1$ &  $3$ & $1.1$  & $1\times 10^{-11}$ & $1/25$ \\
D &  $1$ & $10$ & $1.2$  & $1\times 10^{-11}$ & $1/25$ \\
E &  $1$ &  $3$ & $1.2$  & $2.4\times 10^{-10}$ & $1$ \\
F &  $1$ &  $3$ & $1.2$  & $2.4\times 10^{-12}$ & $1/100$ \\
G &  $1$ &  $0.5$ & $1.2$  & $2.4\times 10^{-12}$ & $1/100$ \\
\enddata
\end{deluxetable}

\begin{description}
\item[] {\it Stellar Rotation:} It is believed that, after the dissipation of the disk, a low-mass PMS star spins up as it contracts towards the zero age main sequence. Such belief of a disk-regulated rotational evolution during the PMS phase is supported by observations of the bimodal distribution of rotation periods \citep{1993A&A...272..176B, 1996AJ....111..283C, 2006ApJ...646..297R}. In samples of PMS stars from the Orion Nebula Cluster and NGC2264, \citet{2007ApJ...671..605C} found that stars with disks possess a distribution of periods peaking around $P \sim 8$~d, while stars without disks presents the peak of the bimodal distribution at $P \sim 2$~d. To explore the extreme values of $P$, we chose periods ranging from $0.5$~d to $10$~d.

\item[] {\it Coronal Temperature:} Over the last years, several observations, such as from the large X-ray surveys Chandra Orion Ultradeep Project \citep{2005ApJS..160..319G} and XMM-Newton Extended Survey of the Taurus Molecular Cloud \citep{2007A&A...468..353G}, showed that the coronal temperature of TTSs can exceed $10^7$~K. It is believed that the bulk of the X-ray emission is coronal and it comes from the material confined inside closed magnetic field structures \citep{2005ApJS..160..401P}. However, the outflowing stellar wind is believed to be much cooler, as estimates presented by \citet{2007ApJ...655..345J} from UV spectra of the CTTS TW Hya and by \citet{2007ApJ...654L..91G} for the CTTS RY Tau. In the present paper, our stellar wind models adopt a hot corona whose temperature at $R_*$ is $T_0=1 \times 10^6$~K. Future works exploring latitudinal dependence of coronal heating, such as a prescribed heating for loops, will be explored.

\item[] {\it Magnetic Field Intensity: }  High mean surface field strengths of up to a few kG have been observed in T Tauri stars \citep[e.g.,][]{2007ApJ...664..975J,2007MNRAS.380.1297D, 2008AJ....136.2286Y}. Although the magnetic field configuration at the surface seems complex, a dipolar component should dominate at larger distances. A dipolar magnitude of few hundreds of G has been detected \citep{2004Ap&SS.292..619V}. In this work, we chose a magnetic field intensity in the range of $200$ -- $1000$~G. 

\item[] {\it Ratio of specific heats:} When detailed heating and cooling mechanisms are unknown, the energy budget of a stellar wind can be described by the use of a heating parameter $\gamma$ (Eq.~\ref{eq:energy_density}). Near the Sun, for instance, due to a large amount of heating (e.g., by turbulence, dissipation of MHD waves, etc.), the solar plasma is considered as a gas with $\gamma \sim 1$ \citep{1988JGR....9314269S}. In T Tauri stars, the heating processes of the corona are poorly known. For simplicity, we parametrize the energy content of the wind of a WTTS by the use of $\gamma$ and we adopt $\gamma =1.1$ -- $1.2$.

\item[] {\it Density:} The range of densities we adopted for the base of the stellar wind is $2.4\times 10^{-12}$ -- $2.4\times 10^{-10}$~g~cm$^{-3}$. This choice of density results in mass-loss rates ranging between $\sim 10^{-9}$ and $8 \times 10^{-8}$~M$_\odot$~yr$^{-1}$.
\end{description}

\section{RESULTS: THE MAGNETIC FIELD CONFIGURATION}
\begin{figure*}
  \includegraphics[scale=0.35,angle=270]{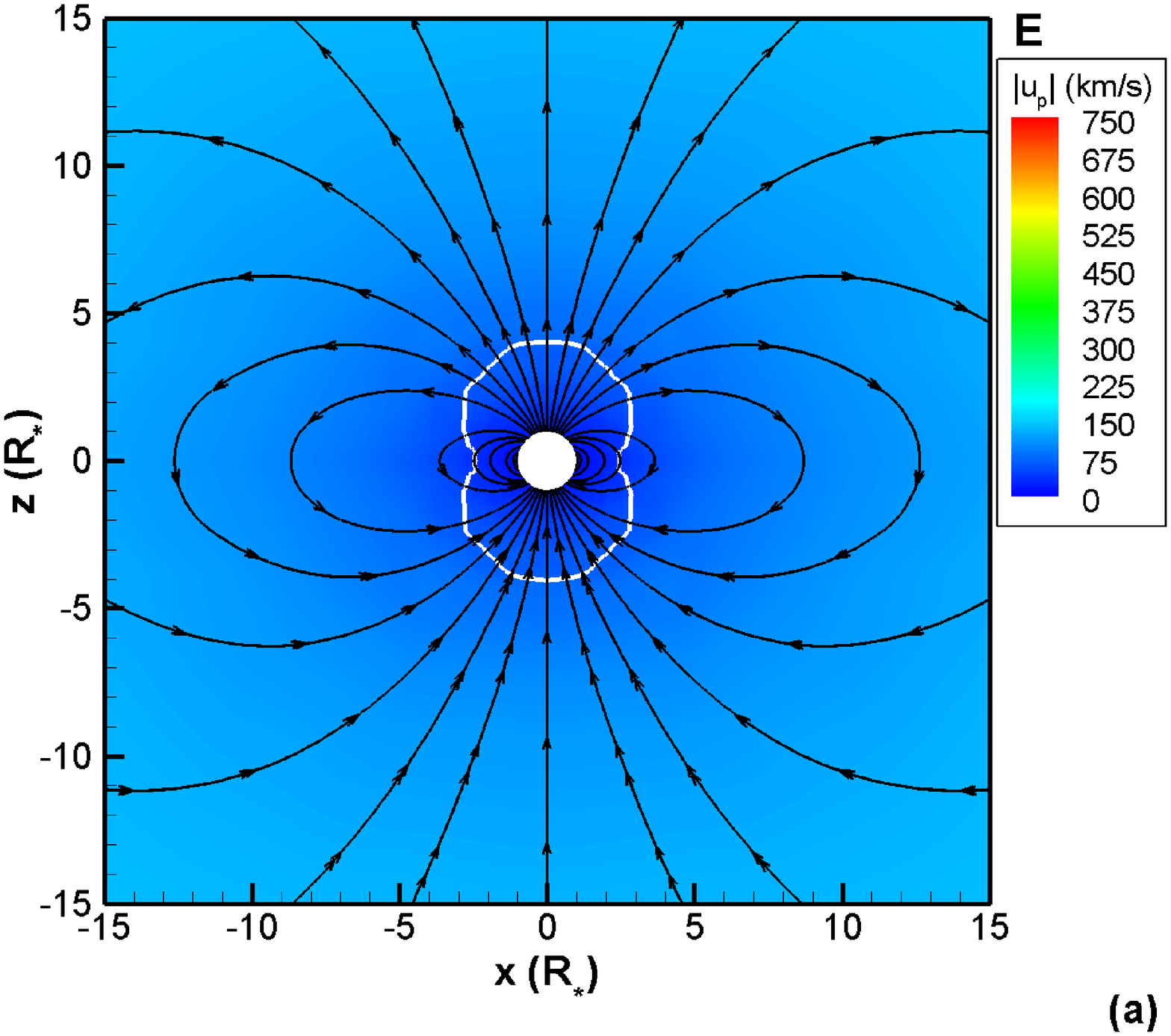}
  \includegraphics[scale=0.35,angle=270]{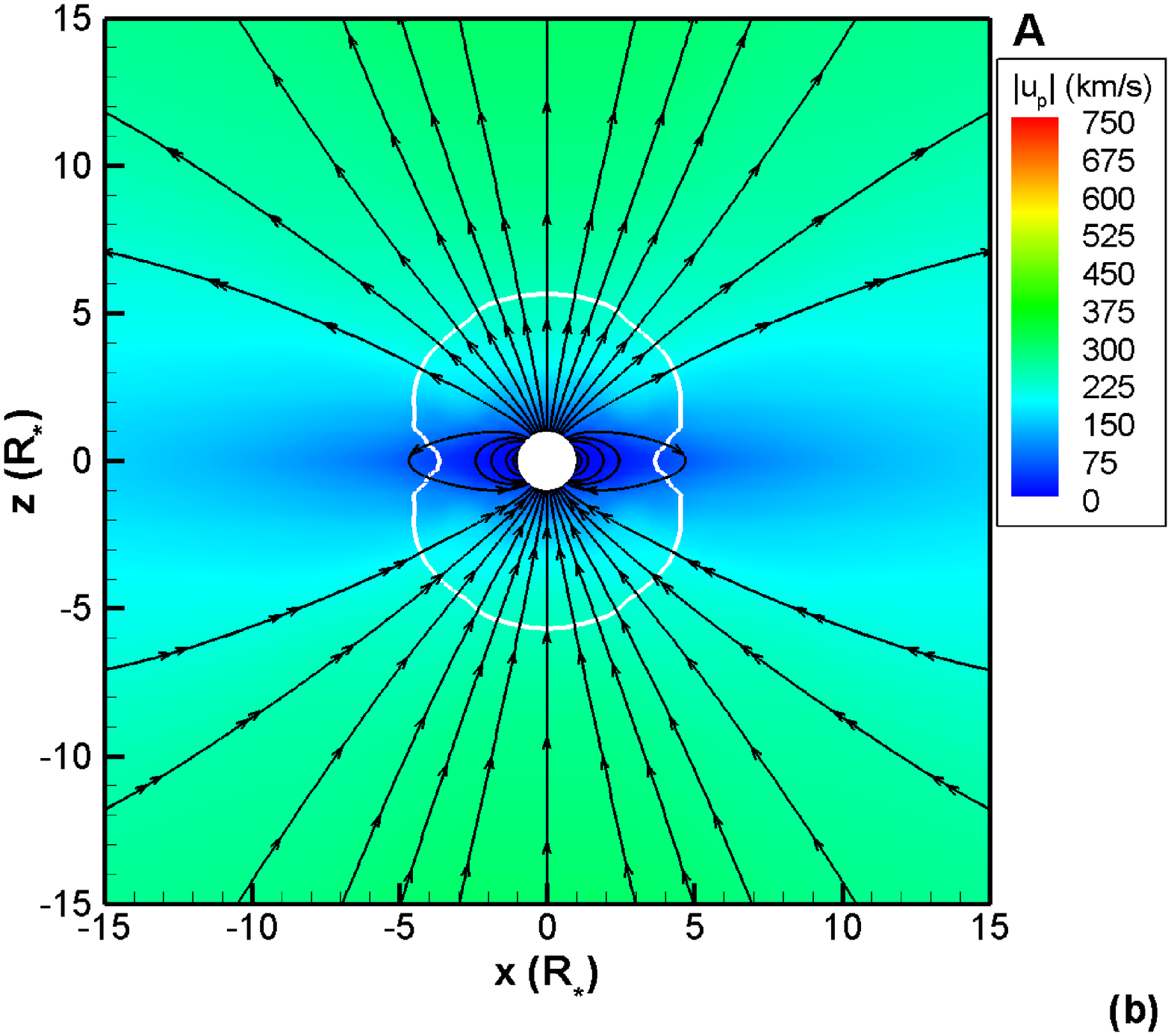}\\
  \includegraphics[scale=0.35,angle=270]{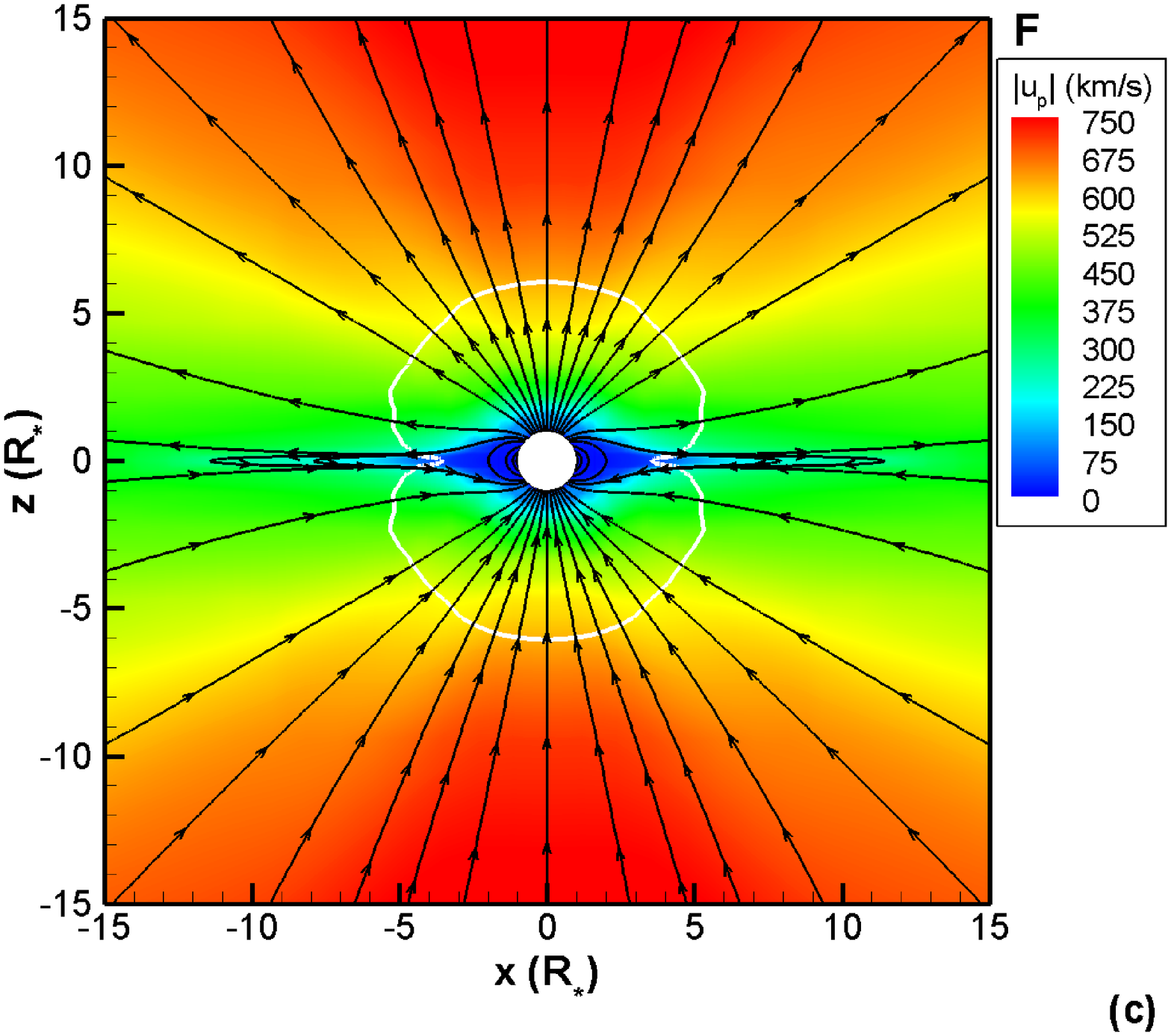}
  \caption{Meridional cuts of poloidal velociy profiles for: (a) case E ($\beta_0=1$), (b) case A ($\beta_0=1/25$), (c) case F ($\beta_0=1/100$). The black lines are streamlines of magnetic field and the white lines represent the Alfv{\' e}n surface. \label{fig:velocity} }
\end{figure*}

In Fig.~\ref{fig:velocity}, we present meridional plots for cases A, E, and F: the plots show contours of the poloidal velocity, the black lines are streamlines of magnetic fields, and the white lines represent the Alfv{\' e}n surface (where the poloidal velocity of the wind equals the local Alfv{\' e}n velocity). The first immediate conclusion is that the configuration of the magnetic field lines is very different in each case. Case E (Fig.~\ref{fig:velocity}a) presents a dipolar-like configuration of the magnetic field lines. In case F (Fig.~\ref{fig:velocity}c), the open-field lines are more collimated in the polar region and in the equatorial region, we observe the formation of streamers. In the intermediate case A (Fig.~\ref{fig:velocity}b), we find a larger region of closed-field lines, coexisting with an open-field line region emanating from high latitudes of the star. What differs between these three cases is the plasma-$\beta$, i.e., the ratio of the thermal to magnetic energy densities. For cases E, A, F,  $\beta_0$ (evaluated at the pole) is respectively $1$, $1/25$, $1/100$.

As discussed in \citet{paper1} in the context of non-rotating main-sequence solar-like stars, the plasma-$\beta$ is a decisive parameter in the acceleration of the wind. When the magnetic energy density is of the same order as the thermal energy density, as in case E (Fig.~\ref{fig:velocity}a), the magnetic configuration remains approximately that of a dipole. Otherwise, when the magnetic energy density dominates over the thermal energy density, the wind is accelerated reaching a situation where the flow ram pressure is able to distort the initial magnetic configuration, as in case F (Fig.~\ref{fig:velocity}c).

In addition to the acceleration of the flow, as a consequence of the Lorentz force, the wind velocity becomes more anisotropic with the decrease of $\beta_0$. Among the cases we analyzed, case E (and also B) presents velocity profiles resembling spherical symmetry and in case F, where $\beta_0=1/100$ (high magnetic energy density), the wind is visibly divided into a region of fast speeds at high latitudes and slow speeds at low latitudes. 

A comparison between cases A and F, where open-field lines are found, shows that: (i) case F shows higher collimation of lines towards the axis of rotation; (ii) the ratio of open to closed field line is larger in case F. As the magneto-centrifugal forces are responsible for the collimation of lines along the axis of rotation, it is expected that a wind more magnetically dominated would present higher levels of collimation (considering same stellar rotation rate). The reason why there are more open magnetic field lines in case F is mainly due to the flow speed: the high wind velocities stretch and open the previously closed-field lines. As the polar wind is faster in case F, it is natural to expect a higher ratio of open to closed field line when compared to case A.

These results also show that longer streamers will be present only if $\beta_0 \ll 1$. The presence of elongated magnetic features has been recently considered to explain observations of recurrent radio flares in binary systems composed of T Tauri stars \citep{2008A&A...480..489M,2008MNRAS.385..708S}. In these systems, the elongated magnetic features belonging to both stars could overlap in certain orbital phases, resulting in a modification of the magnetic configuration (reconnection). This picture could account for giant radio flares, emitted when the binary system is near periastron. Taking for example the system under consideration in \citet{2008A&A...480..489M}, the existence of elongated magnetic features indicates that $\beta_0 \lesssim 0.01$. Adopting $B_0 \sim 2600$~G and $T_0\sim 1$~MK, it results in $n_0 T_0 \lesssim 2 \times 10^{19}$~K~cm$^{-3}$, or $\rho_0 \lesssim 3.3 \times 10^{-11} \mu$~g~cm$^{-3}$, where $n_0$ is the number density at $R_*$ and $\mu$ is the mean molecular weight of the wind particles. Therefore, the observations of radio flares can help us constrain the coronal density in these stars.

PMS stars exhibit X-ray flares occuring in star-sized, small structures, often interpreted as scaled-up solar-like coronal activities \citep{2005ApJS..160..401P}. In addition, intense X-ray flares are also observed. \citet{2005ApJS..160..469F} analyzed these intense X-ray flare decays to model physical parameters of the flaring structure, finding that these large flares take place in very long magnetic structures, extending out to several stellar radii. \citet{2008ApJ...688..437G} also argued that in fast rotating stars these powerful X-ray flares extend beyond the corotation radius, where centrifugal forces dominate over gravity. They conclude that the flares arise in traditional solar-type magnetic loops, where both footpoints are anchored in the stellar surface. \citet{2005MNRAS.361.1173J} developed a model for magnetic loops in rapidly rotating main-sequence stars that explains the existence of magnetic prominences that can extend out to heights of $2$-$5~R_*$, above the corotation radius, in equilibrium with the open field region. In our simulations, we observed an extended, slender closed field region in case F. In this case, the extension of the streamer is assured up to an equatorial radius of $\sim 13~R_*$. Above this point, the grid changes resolution, causing a local numerical reconnection (in principle, the loop could extend farther out). At the equatorial plane, the corotation radius is given by
\begin{equation}
r_{\rm co} = \left( \frac{G M_\star P}{4 \pi^2} \right)^{1/3} = 4.2 R_\odot \left( \frac{P}{1 {\rm ~d}} \right)^{2/3}\left( \frac{M}{M_\odot} \right)^{1/3} \, ,
\end{equation}
which for the TTS simulated here is $r_{\rm co}/R_*  = 1.95 P_d^{2/3}$, where $P_d$ is the stellar rotation period measured in days. For $P=3$~d, $r_{\rm co} \simeq 4 R_*$. This shows that the elongated streamers, as obtained in case F, is stable to at least $\sim 3$ times the corotation radius. 

\begin{figure*}
  \includegraphics[scale=0.35,angle=270]{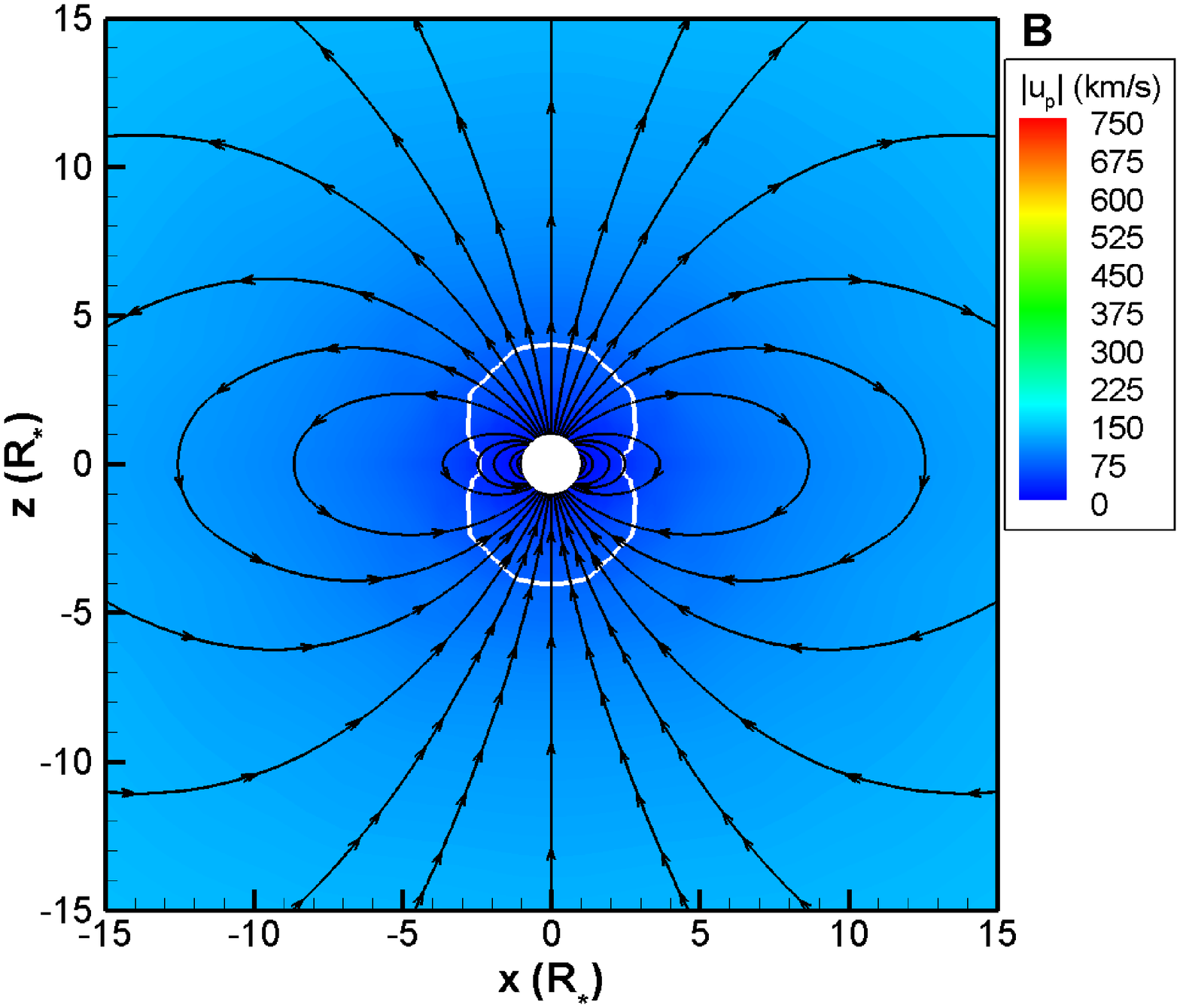}
  \includegraphics[scale=0.35,angle=270]{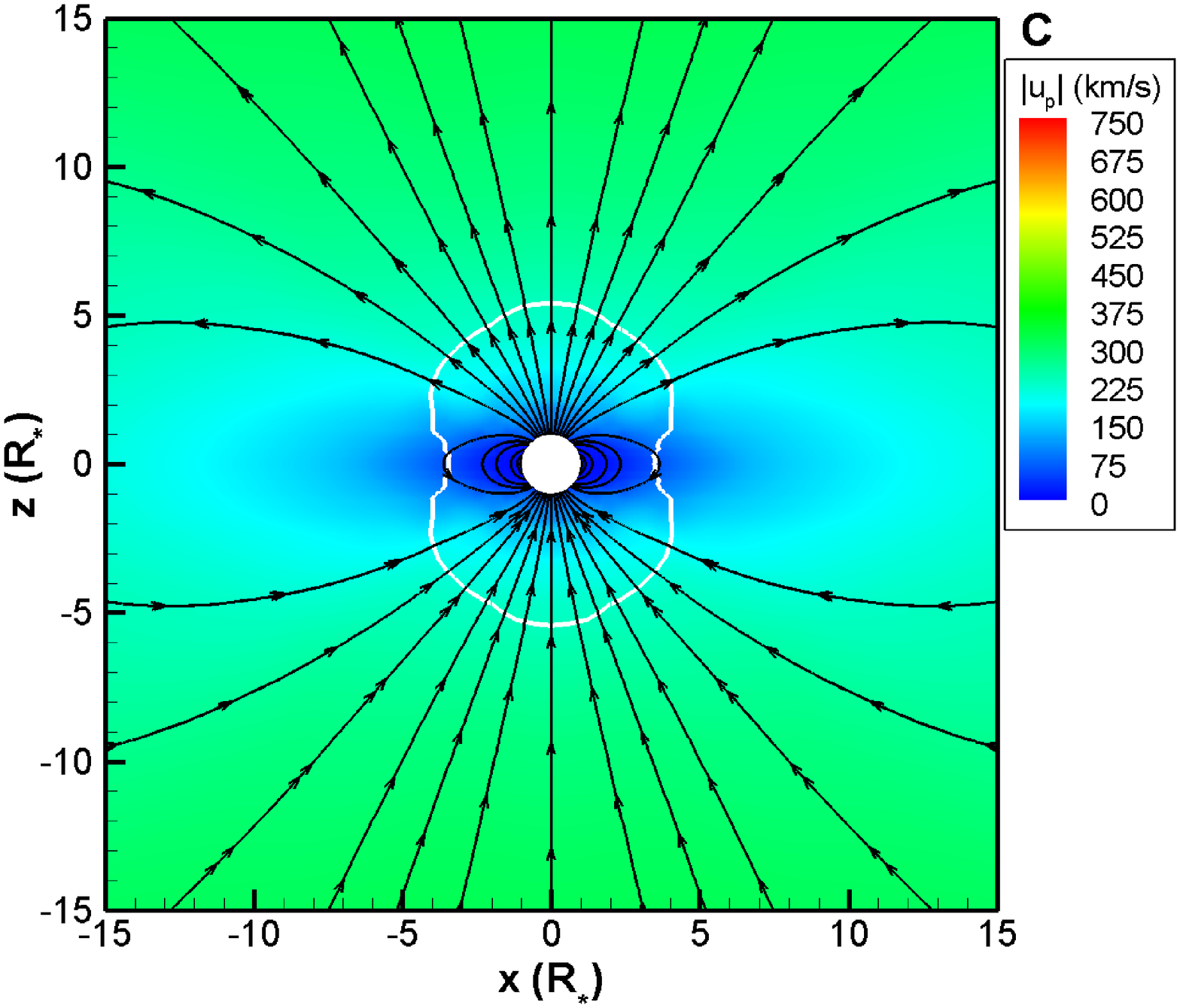} \\
  \includegraphics[scale=0.35,angle=270]{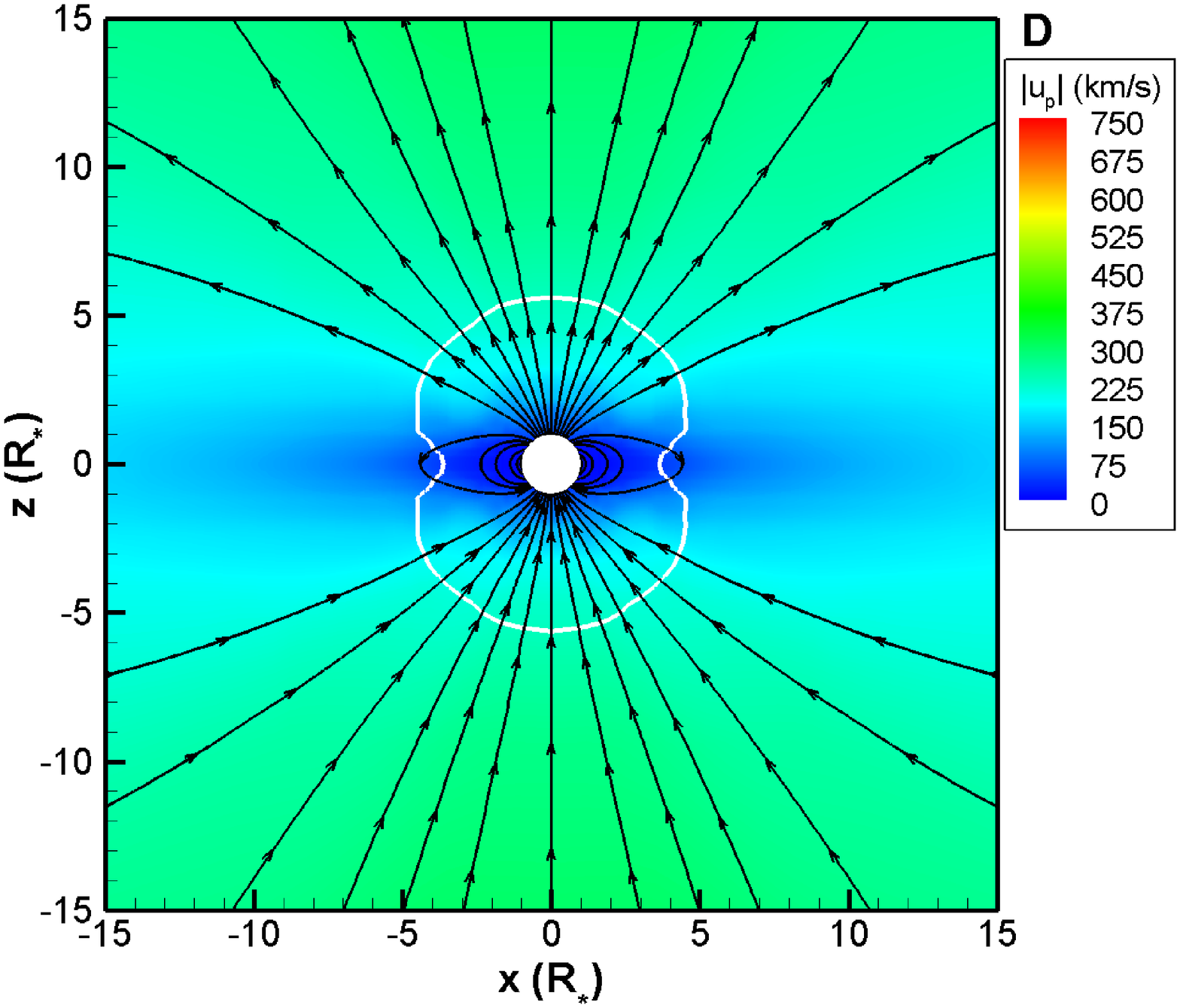} 
  \includegraphics[scale=0.35,angle=270]{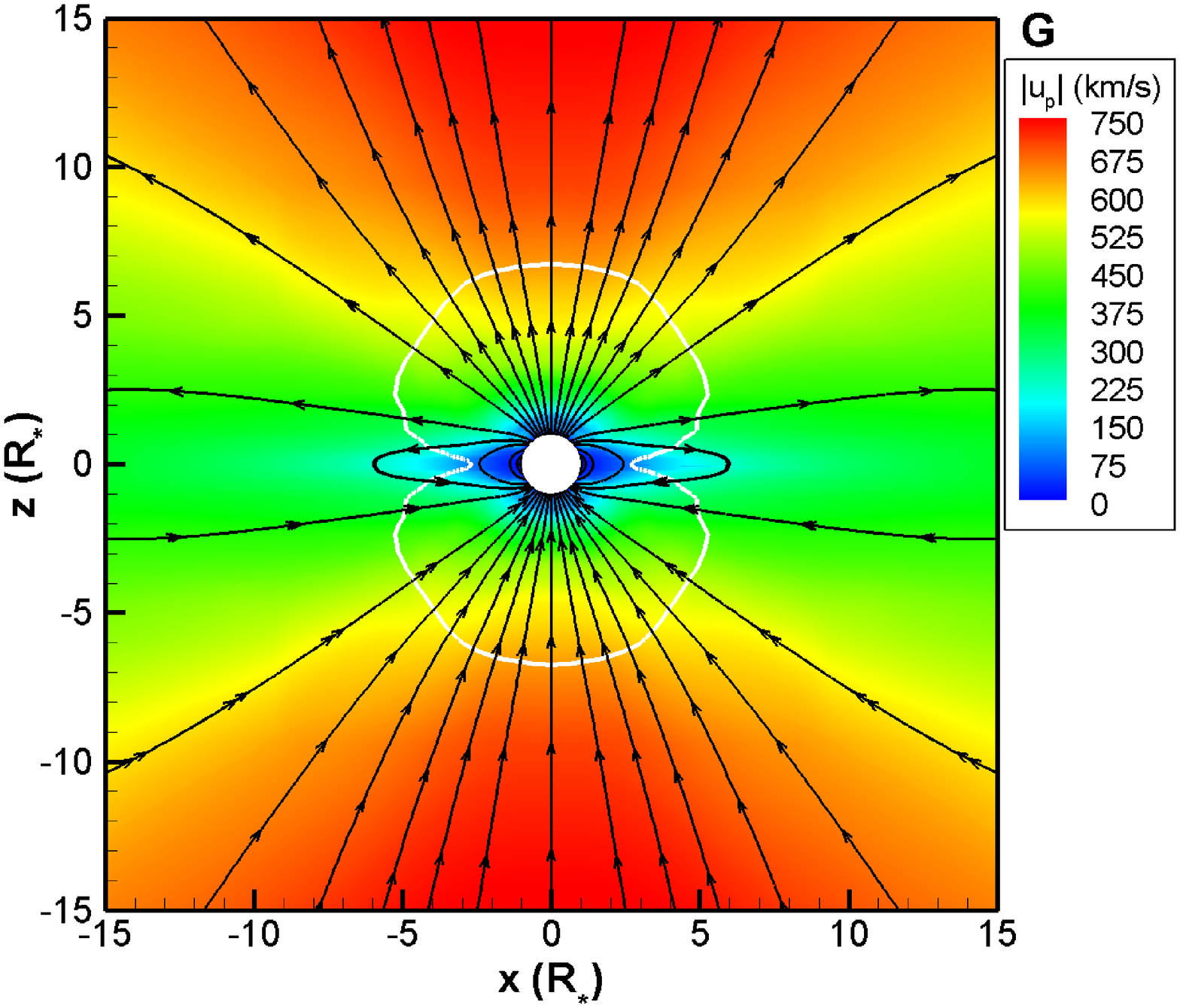}
  \caption{Same as Fig.~\ref{fig:velocity} for the remaining cases.  \label{fig:velocity2}}
\end{figure*}

Figure~\ref{fig:velocity2} shows the steady-state solutions for the remaining of the simulations presented in Table~\ref{table:4}. We enumerate the following conclusions: ({\it i}) We show that only a change in the rotation of the star does not provide a significant change in the poloidal velocity profile of the system (compare cases A and D, F and G). It changes, however, how the magnetic field lines are twisted: the faster is the rotation of the star, the larger is the $\varphi$-component of the magnetic field. ({\it ii}) The heating parameter $\gamma$ influences the thermal acceleration of the wind (compare cases A and C), being more accelerated in the case where the parametrized energy input is higher, i.e., for low $\gamma$. As the wind characteristics are different, the resultant interaction with the magnetic field is different: case A (high $\gamma$) presents a larger ratio of open to close lines than case C (case C has a larger closed-field line region). ({\it iii}) Cases B and E present the same characteristics, as they possess the same $\beta_0$ \citep{paper1}. Furthermore, because $\beta_0=1$, there is an equipartition of thermal and magnetic energy densities at the base of the corona, which results in lower velocities and field lines that are merely modified in relation to the initial configuration, resembling a dipole. ({\it iv}) Case G represents a star with a high rotation rate. Because of this, among the cases we analyzed, it is the case where the open field lines are more collimated towards the axis of rotation. Comparing it to case F, where the rotation rate is 6 times smaller, it can also be seen that the higher the rotation rate is, the less elongated are the closed field lines.

\begin{figure}
  \includegraphics[scale=0.37,angle=270]{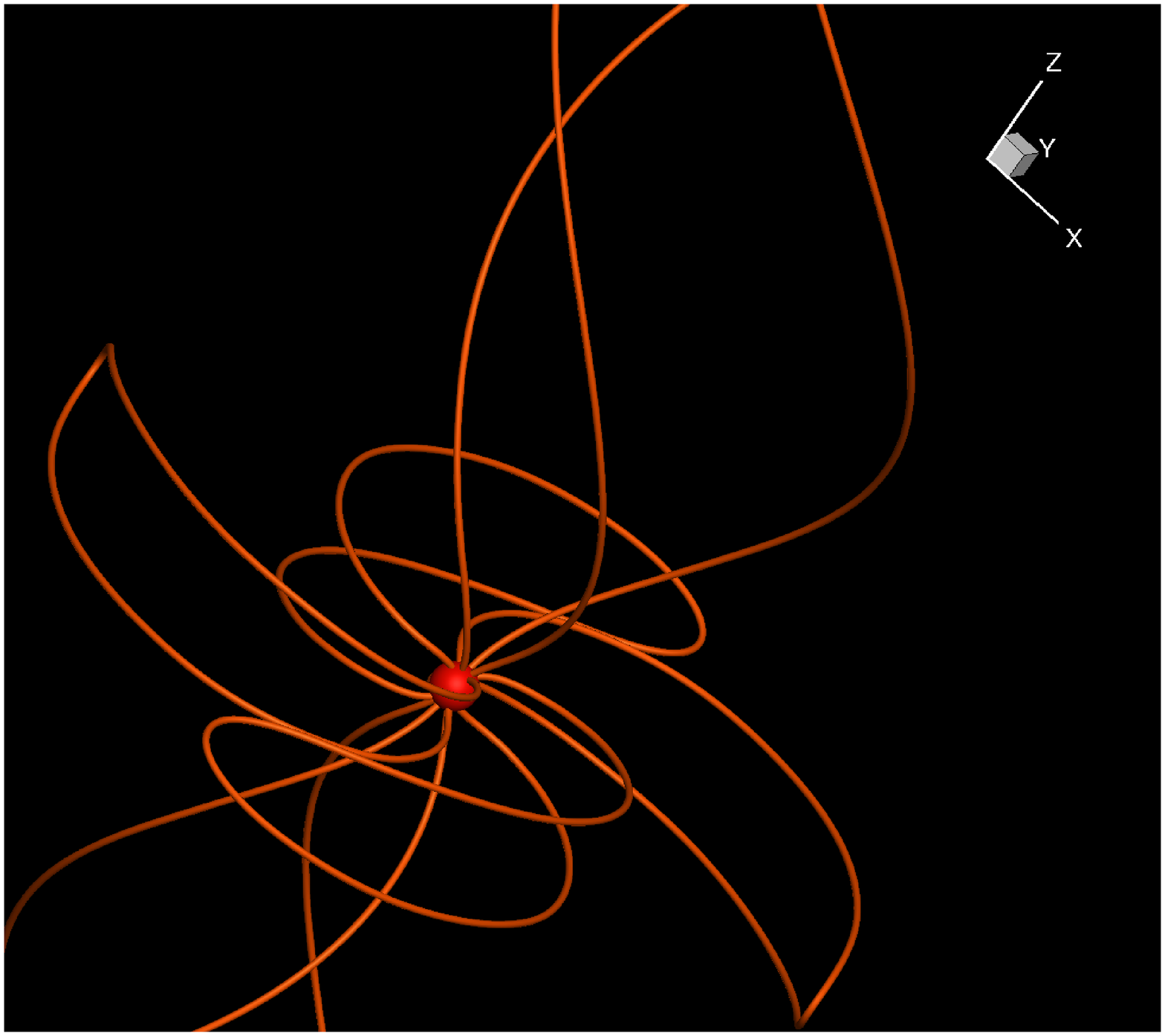}%
  \caption{Three-dimensional view of the inner most region of case G. The lines plotted were selected as to outline the twisting of both closed and open magnetic lines anchored on the surface of the star.  \label{fig:3d-lines} }
\end{figure}

The Alfv{\' e}n surface plays an essencial role in the determination of the magnetic field configuration. In magneto-centrifugal winds, both $B_\varphi/B_r$ and $B_\varphi/B_\theta$ are expected to be small inside the Alfv{\' e}n surface. Beyond the Alfv{\' e}n surface, the $r$ and $\theta$ components of the magnetic field decay faster than the $\varphi$ component, causing the field lines to be twisted. Therefore, helmet streamers anchored on rotating stars are not radial. In Fig.~\ref{fig:3d-lines}, we plot a 3D view of the region close to the star. As can be seen, the picture of approximately radial streamers, as observed in the Sun or in slowly rotating stars, is not anymore verified in WTTSs with high rotation rates: both open and closed lines become twisted due to the stellar rotation.

\citet{1988ApJ...335..940A} presented an analytical expression for the calculation of the Alfv{\' e}n surface, considering that (i) the wind should have spherical mass-loss; (ii) by the time the wind is crossing the Alfv{\' e}n surface, it has already achieved the terminal velocity; (iii) the field is dipolar all the way to the Alfv{\' e}n surface. As can be seen from our results, when the self-consistent interaction of the field lines and the wind is taken into account, these hypotheses may not be verified any longer.

\citet{2008A&A...480..489M} computed the Alfv{\' e}n radius of the wind of each of the stars in the binary system V 773 Tau A, where they observed helmet streamers, using \citet{1988ApJ...335..940A}'s analytical expression. They ruled out the twisting of magnetic lines of the helmet streamer, because in their estimate they found a large Alfv{\' e}n surface with radius of $\sim 60~R_*$. According to our model, the Alfv{\' e}n surface of each individual stellar wind is located much closer to the star, and is probably even smaller than the minimum separation of the stars in the binary system V 773 Tau A ($\sim 30~R_*$), suggesting that the interacting streamers are probably twisted, as illustrated in Fig.~\ref{fig:3d-lines}.

\section{STELLAR WIND EFFECTS ON PLANET MIGRATION}
In the previous section, we investigated the magnetic configuration of the wind of WTTSs. This configuration arises naturally from the interplay between the outflow and the field, i.e, the magnetic field is not restricted in our simulations. Several models assume a fixed magnetic field topology, as is the case of the Weber \& Davis (WD) model. In this section, we investigate the effects our more realistic wind model will have on the migration of a giant planet. 

According to the current favored planetary formation theory, the core of a giant planet forms far away from the central star, beyond the snow line ($\gtrsim 5$~AU, where the disk temperature falls below the condensation temperature of water) by coalescence of planetesimals \citep{1996Icar..124...62P}. When the solid core is sufficiently massive, it captures the surrounding gas of the accretion disk to form the atmosphere of a giant planet \citep{2005A&A...433..247P}. However, observations show the existence of planets orbiting solar-like stars at very close distances ($\lesssim 0.1$~AU). For this reason, these giant planets are often referred as hot-Jupiters. A possible explanation for this inconsistence is that the giant planet may have formed several AU from the central star, and later migrated inward.  The interaction of the protoplanet with the disk wherein it was formed is a probable process that may drive this migration \citep{1996Natur.380..606L}. 

The inward migration is expected to cease when the planet is in the inner magnetospheric cavity, a cleared region between the stellar surface and the inner region of the accretion disk \citep{1996Natur.380..606L, 1998ApJ...500..428T,2007A&A...463..775P}. In recent numerical simulations, \citet{2006ApJ...645L..73R} found that for a stellar dipolar field that is not highly misaligned with the spin axis of the star neither almost aligned so that 3D instabilities cannot fill the magnetospheric cavity with matter, the formation of the cleared region is assured. As in the cavity there is insufficient external material to remove the planet's angular momentum, the giant planet radial motions halt; the planet cannot grow further. In addition to this mechanism, others have been proposed to explain why the planet migration is inhibited all the way to the star \citep{1996Natur.380..606L, 2008Ap&SS.313..351F}. 

In this section, in particular, we will investigate the action of magnetic torques from the stellar wind acting on the planet.

Considering that a T Tauri star has magnetic fields and rotation rates that are several times larger than in the Sun, Lovelace et al. (2008, LRB08 from now on) studied the torque that a magneto-centrifugally driven wind would cause on a close-in giant planet. They adopted the model of WD for the stellar wind and evaluated the change of the planet's angular momentum due to an azimuthal drag force exerted by the wind on the planet. Considering a solar-mass star with $2~R_\odot$, $P \sim 3 $ -- $5$~d, and $B_0 \sim 1$ -- $3$~kG, they found a time-scale of $2$ -- $20~$Myr for a giant planet like Jupiter, orbiting the equatorial plane of the star, to have considerable radial motions. 

Following \citet{2008MNRAS.389.1233L}, we estimate this time-scale by adopting a more realistic stellar wind, product of our simulations. Consider a planet of mass $M_p$, orbiting very close to the star at a distance $r_p$. The angular momentum of the planet is then $L_p =M_p v_K r_p$, where $v_K = (G M_\star/r_p)^{1/2}$ is the azimuthal velocity of the planet, assumed Keplerian. Therefore, a change in the planet's angular momentum leads to
\begin{equation}\label{eq.torque.planet}
\left| \frac{d L_p}{dt} \right| \simeq \frac12 M_p v_K \frac{d r_p}{dt} \simeq \frac12 M_p v_K \frac{r_p}{\tau_w}\, ,
\end{equation}
where $\tau_w$ is the time-scale for the stellar wind drag change significantly the planet's orbit \citep{1996Natur.380..606L}. Another way to express the torque (Eq.~\ref{eq.torque.planet}) is to multiply the force the stellar wind will exert on the planet by the distance $r_p$ \citep{2008MNRAS.389.1233L} 
\begin{equation}\label{eq.torque.planet2}
\left| \frac{d L_p}{dt} \right| \simeq \ (P_{\rm ram} A_{\rm eff}) r_p\, ,
\end{equation}
where $P_{\rm ram}$ is the ram pressure the wind exerts on the planet
\begin{equation}\label{eq.pram}
P_{\rm ram} = \frac12 \rho (u_\varphi - v_K)^2 + \frac{1}{4 \pi} B^2\, ,
\end{equation}
and $A_{\rm eff}$ is the cross-section of the planet that intercepts the stellar wind. In order to take into account the effects that the magnetic field of the planet could have in the drag force of the stellar wind, \citet{2008MNRAS.389.1233L} defined an effective cross-section of the planet as
\begin{equation}\label{eq.aeff}
A_{\rm eff} = {\rm max} \left[ \pi R_p^2 \left( \frac{B_p^2/4\pi}{P_{\rm ram}} \right)^{1/3}, \pi R_p^2 \right]\, ,
\end{equation}
where $B_p$ is the surface magnetic field intensity at the pole of the planet and $R_p$ is its radius. The planet is assumed to have a dipolar-field configuration, and planetary mass loss is not considered. From Eqs.~(\ref{eq.torque.planet}) and (\ref{eq.torque.planet2}), we find
\begin{equation}\label{eq.time-scale}
{\tau_w} \simeq \frac12  \frac{M_p v_K}{P_{\rm ram} A_{\rm eff}} \, .
\end{equation}
A large $A_{\rm eff}$ means that the cross-section of the planet intercepting the wind is large, and thus, a more efficient drag mechanism (i.e., low $\tau_w$) is expected. If the wind ram pressure is large, the dragging mechanism is also expected to be important. However, a high $P_{\rm ram}$ also implies in a small $A_{\rm eff}$ [see Eq.~(\ref{eq.aeff})].

The planet is supposed to be in the corotation radius, if the system is in a tidal equilibrium state (i.e., synchronized with the stellar rotation). However, \citet{2009ApJ...692L...9L} have shown that in the majority of the observed transiting extra-solar planets, the star's rotation is not synchronous. Due to this reason, we compute $\tau_w$ for a range of equatorial radial distances. 

\begin{figure}
  \includegraphics[scale=0.35,angle=270]{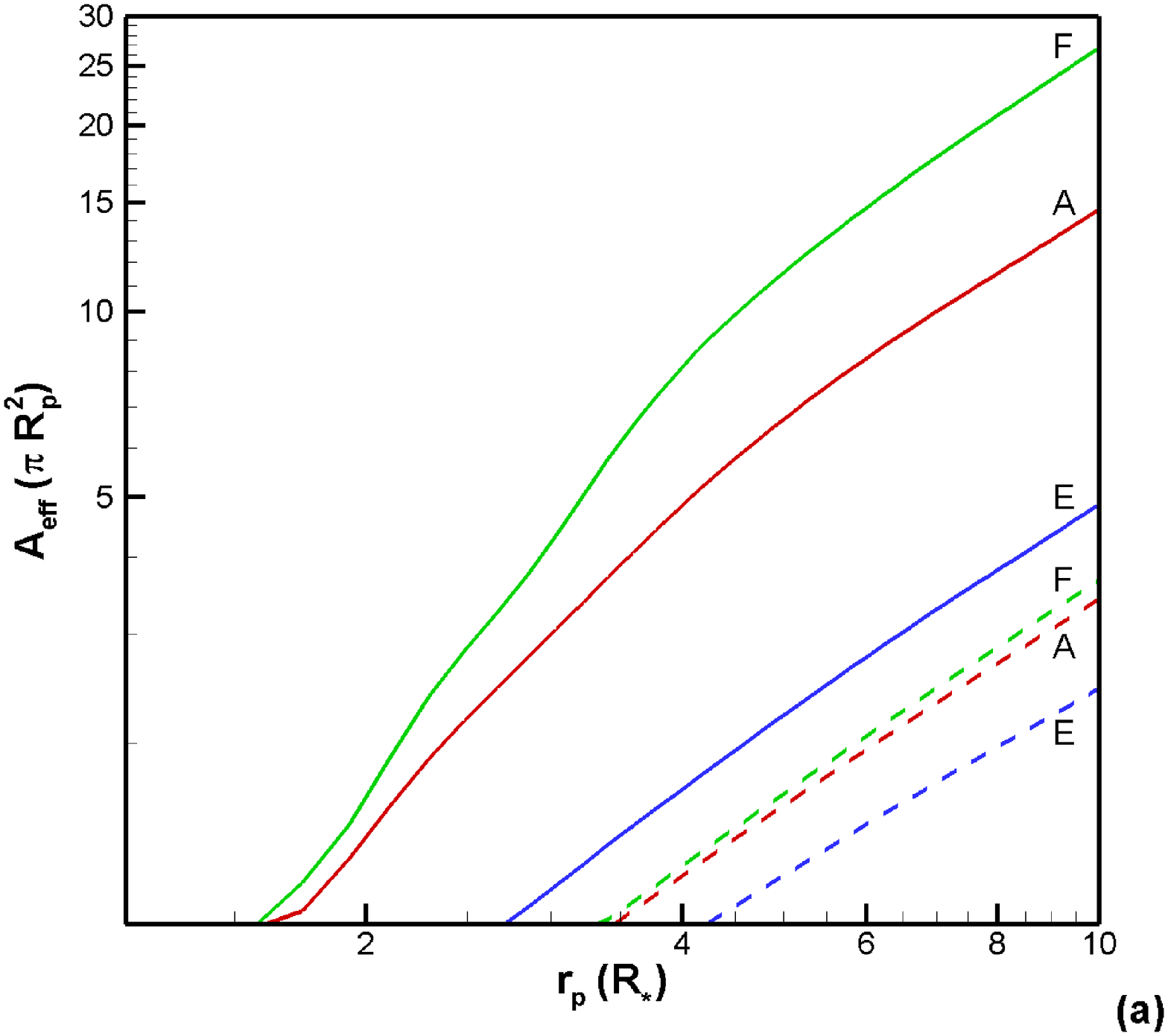}\\%
  \includegraphics[scale=0.35,angle=270]{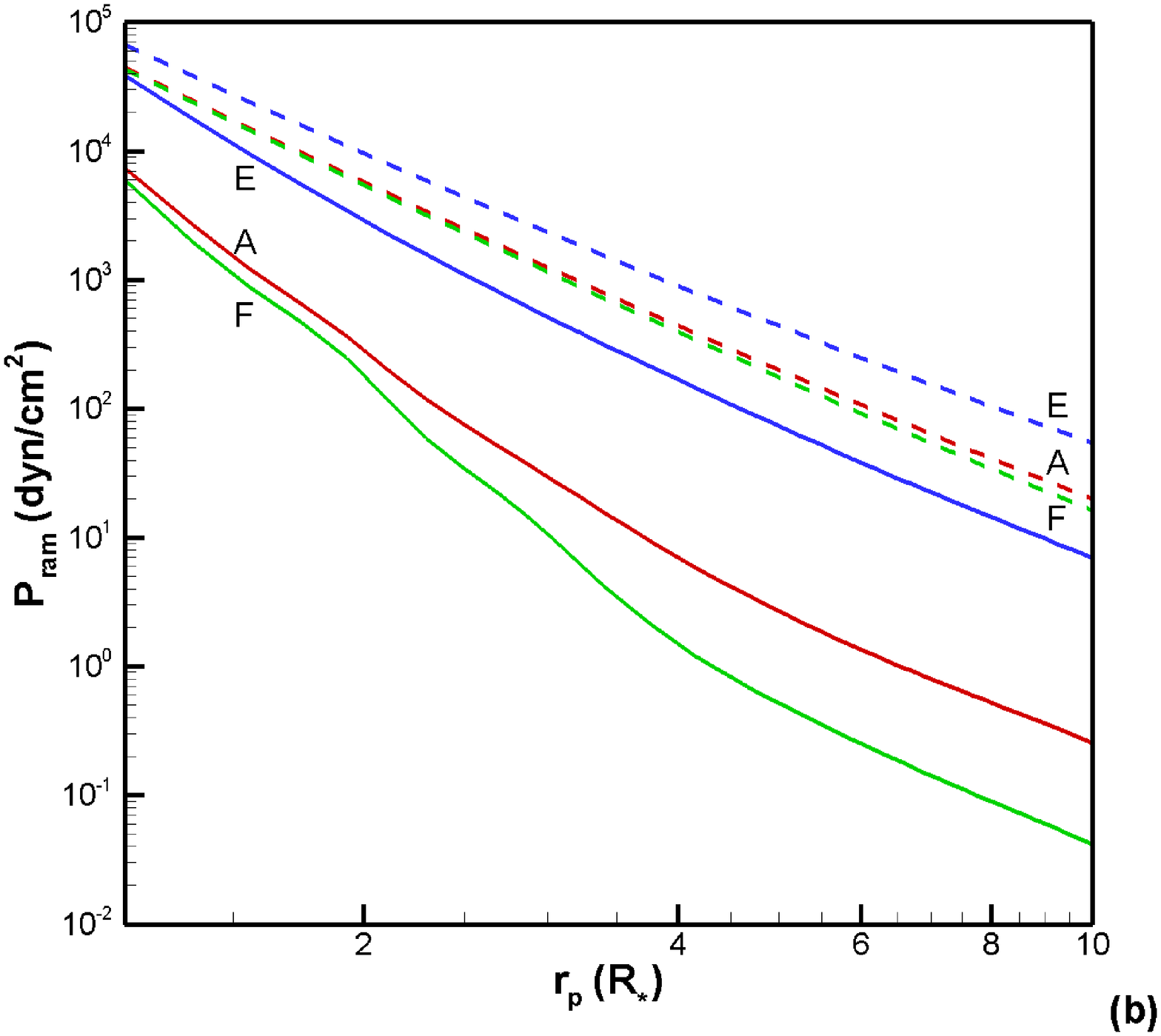}\\%
  \includegraphics[scale=0.35,angle=270]{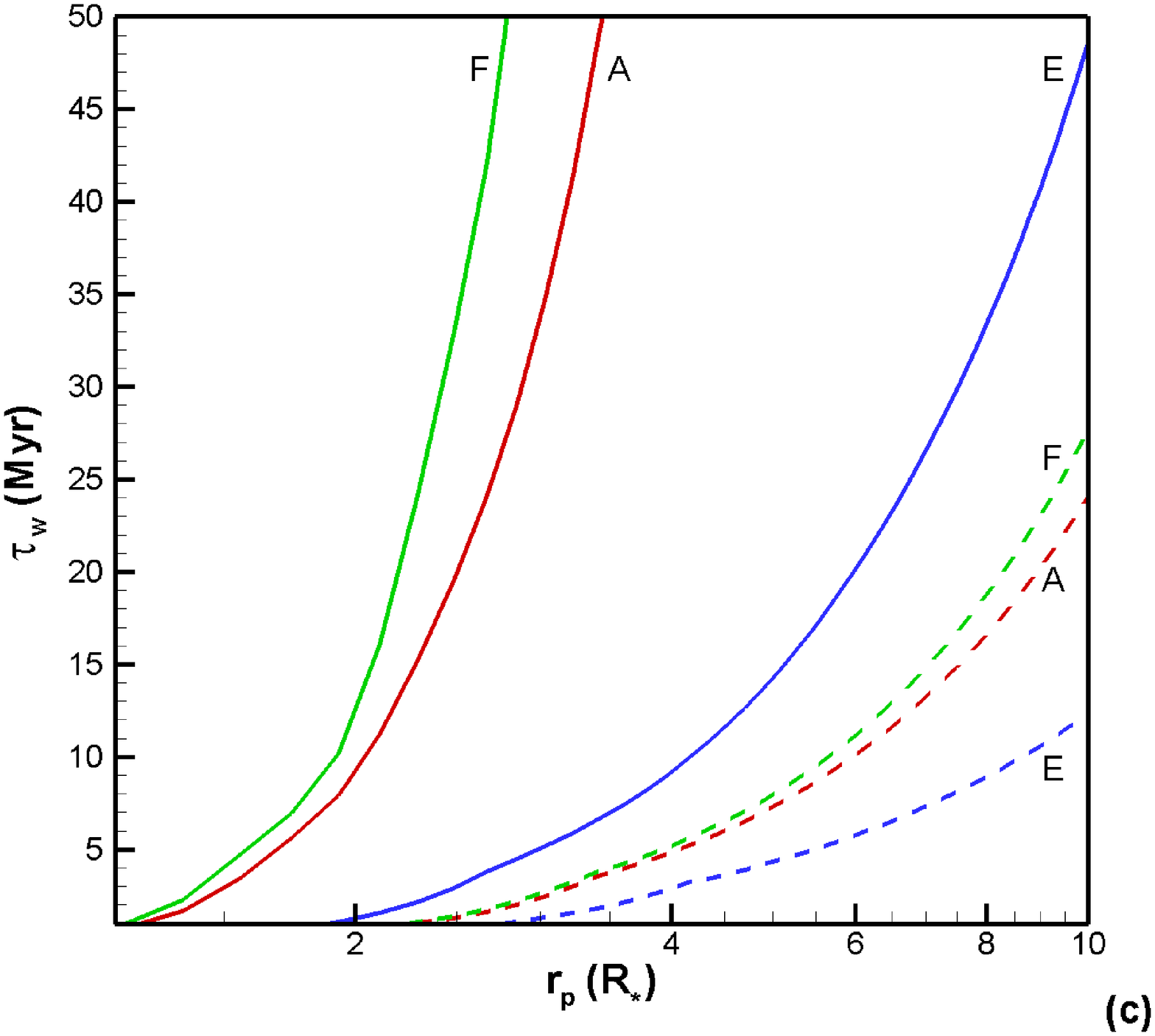}%
  \caption{Comparison of: a) The effective cross-section of the planet that intercepts the stellar wind. b) The wind ram pressure. c) The time-scale for the stellar wind drag change significantly the planet's orbit. We use the same parameters as in cases A (red lines), E (blue lines), F (green lines): solid lines are the results for our stellar wind model, and dashed lines for the WD model. \label{fig:Tw} }
\end{figure}

Considering the stellar winds from the simulations A, E, and F, we calculated $\tau_w$ at the equatorial plane by assuming that the giant planet has the same mass and radius as Jupiter, and that the planet's magnetic field is $B_p =100~$G \citep[same parameters adopted by][]{2008MNRAS.389.1233L}. In order to perform the comparison of our results with the WD model, we evaluate Eq.~(\ref{eq.time-scale}) for both models. We thus compute the velocity and magnetic field profiles of the stellar wind using the WD model. We assume the same parameters at the base of the wind as used in our simulations A, E, F, with the difference that $B_0$ is the magnetic field intensity at the pole in our model, and in the WD model it is the radial field calculated at the surface. We do the same procedure for the more realistic winds we simulate, i.e., we use the results of the simulations for ${\bf u}$, ${\bf B}$, and $\rho$ and evaluate ${\tau_w}$ using Eqs. (\ref{eq.pram}) to (\ref{eq.time-scale}). These results are shown in Fig.~\ref{fig:Tw}, where the solid lines represent the results obtained using our model and the dashed lines for the WD model.

\begin{figure}
  \includegraphics[scale=0.35,angle=270]{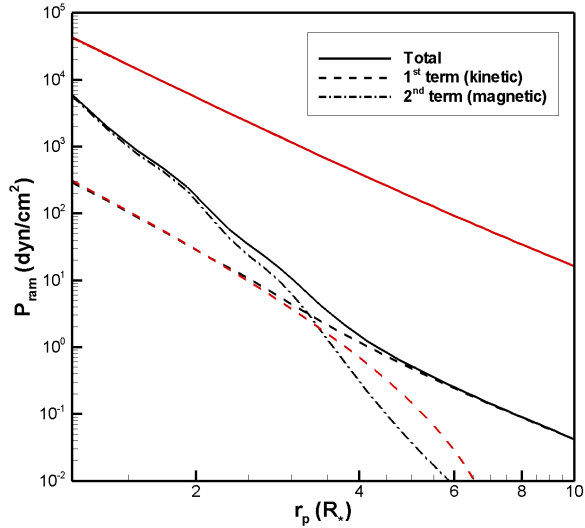}%
  \caption{Relative importance of the first (dashed lines) and second (dot-dashed lines) terms in the RHS of Eq.~(\ref{eq.pram}) calculated for case F using the results of our simulations (black) and using the WD model (red). Solid lines are the total ram pressures, i.e., the sum of first and second terms. \label{fig:pram} }
\end{figure}

Figures~\ref{fig:Tw}a and \ref{fig:Tw}b show, respectively, the effective cross-section of the planet and the ram pressure exerted on the planet for the region inside $\sim 0.1~$AU (i.e., $\lesssim 10~R_*$). It can be seen that when $P_{\rm ram}$ is large, $A_{\rm eff}$ is small, and vice-versa. In addition, $A_{\rm eff}$ ($P_{\rm ram}$) calculated from our stellar wind model is larger (smaller) than if it were calculated using the WD model. This is because, for the WD model, $P_{\rm ram}$ is verified to be essentially magnetic, i.e., the first term of the right-hand side (RHS) of Eq.~(\ref{eq.pram}) is negligible compared to the second one, while for our stellar wind model, both terms are significant. In addition, for both models, the first term of the RHS of Eq.~(\ref{eq.pram}) are of the same order. The difference thus comes from the magnetic term of Eq.~(\ref{eq.pram}), that for the WD model is larger than for our model. This behavior is illustrated in Fig.~\ref{fig:pram}, for case F.

This results in larger time-scales as compared to the WD model (Fig.~\ref{fig:Tw}c), which implies that the more realistic winds simulated in the present paper are not expected to have significant influence on hot-Jupiter migrations. The time-scales we estimate (e.g., $\gtrsim 50$~Myr for cases A and F, for $r_p \gtrsim 3~R_*$) are much larger than the ones estimated by \citet{2008MNRAS.389.1233L} ($\sim 2 - 20$~Myr). Other effects, such as an increase in the base density or magnetic field intensity, can reduce $\tau_w$. Case E presents the highest base density, while case F presents the lowest one among our simulations. From Fig.~\ref{fig:Tw}c, we note that an increase in $\rho_0$ reduces $\tau_w$. We expect that for $\rho_0>2.4\times 10^{-10}$~g~cm$^{-3}$ with the remaining parameters as adopted in cases A, E and F, the effects of the migration will be more significant. A similar trend, i.e., a decrease in $\tau_w$, is also expected if one increases $B_0$. For example, cases A, E and F assume $B_0=1$~kG. If such value were increased, it would result in a larger ram pressure exerted by the wind on the planet. This would cause a more efficient drag and ultimately it would lower $\tau_w$, and perhaps it would even become smaller than $20$~Myr in the region where $r_p \lesssim 10~R_*$. In this case, the more realistic winds could have a more significant influence on the giant planet migration, though it would still have $\tau_w$ much larger than the one computed with the WD model.

According to \citet{2008MNRAS.389.1233L}, the magnetic winds are expected to be important for planet migration for the case of a multipolar magnetic field rather than a dipole field. In the present paper, we extend this statement to the case of the multi-field component coronae, as the ones analyzed previously, where open and closed field lines coexist. Our model suggests that the stellar wind of these modified-dipole coronae are not expected to have significant influence on hot-Jupiters migration.

If the wind azimuthal velocity $u_\varphi$ is greater than the planet Keplerian velocity, \citet{2008MNRAS.389.1233L} showed that the stellar wind causes the planet to move outwards. In all the simulations we presented, this is the case only for case G, for $r_p \gtrsim  1.26~R_*$. For all the other cases, the wind would act as to push the planet inwards, and the closest the planet is from the star, the stronger/faster would be this dragging.

An aspect that is not investigated in the present paper is when the surface dipole is tilted with respect to the stellar rotation axis. The wind ram pressure is an interplay between magnetic ($\propto B^2$) and kinetic terms ($\propto \rho [u_\varphi-v_K]^2$). In the tilted case, a planet orbiting the rotation equator would be subjected to high magnetic field intensities (since $|{\bf B}|$ is maximum at the magnetic pole and minimum at the magnetic equator). However, it is unclear the latitudinal dependence of $u_\varphi$: in the aligned case, $u_\varphi$ is maximum at the equator; in the misaligned case and in a highly magnetized ambient, the stellar magnetic field is expected to channel the rotating outflow. Hence, by a superposition of the magnetic and kinematic effects, it could be possible that  $\tau_w$ becomes smaller in a tilted configuration. A future work will explore this scenario.

Another interesting configuration is when the orbital axis of the planet is inclined with respect to the spin axis of the star, as it is believed to occur, e.g., in the extra-solar planets HD80606b \citep{2009A&A...498L...5M} and XO-3b \citep{2008A&A...488..763H, 2009ApJ...700..302}. In this configuration, the planet would interact with a stellar field strength that is larger at the position of the planet than at the magnetic equator. This could have an effect on $\tau_w$ that could reduce it.

\section{DISCUSSION AND CONCLUSION}
Recent works indicate the presence of complex magnetic structure, as the one observed in the Sun, on T Tauri stars. This structure is certainly influencing the winds of these stars. WTTS offers a tool for studying the magnetic behavior of low-mass pre-main-sequence stars with the benefit that it is no longer strongly influenced by the presence of an accretion disk. As a consequence, the study of WTTS may eventually have applications on theories of CTTSs and also clues on the physical conditions for planetary formation and/or migration.

In the present work, we showed 3D MHD numerical simulations of magnetized stellar winds in rotating T Tauri stars. With the parameters we selected, we scan the possible values of rotation period ($0.5$~d to $10$~d), magnetic field intensities ($200$~G to $1$~kG), and surface density ($2.4 \times 10^{-12}$ to $2.4 \times10^{-10}$~g~cm$^{-3}$). We show that the plasma-$\beta$ parameter is a crucial parameter in the acceleration of the wind and in shaping the magnetic field lines. If the surface magnetic and thermal energy densities are equal at the pole, i.e. $\beta_0=1$, we showed that: (i) the wind presents approximately spherical symmetry; (ii) the magnetic field lines are not significantly distorted from the dipolar field we assume as initial condition of the simulation; (iii) the wind velocity does not differ considerably from the hydrodynamical case. On the other hand, if the magnetic energy density is much greater than the thermal one, i.e. $\beta_0 \ll 1$, we observed departure of spherical symmetry. The wind shows two different components of the velocity, being faster at high latitudes and slower at low latitudes. In addition, the field lines are not anymore dipolar: the high velocity of the flow is able to stretch and open the field lines at high latitudes, while in low latitudes we observe elongated magnetic features. This indicates that streamer structures should be present in WTTSs with $\beta_0 \ll 1$.

In a different context, \citet{2002ApJ...576..413U} defined a similar parameter that is used as an indicator of the effectiveness of magnetic fields in channeling outflow from hot supergiant stars. The ``magnetic confinement parameter'', defined by them as the ratio between the magnetic and kinetic wind energy densities, operates in a similar fashion as $\beta_0$ operates in our work: it identifies whether the magnetic field will play a significant role in the acceleration of the wind. However, both in the present work and in \citet{paper1}, the thermal energy of the (coronal) wind is more significant then the kinetic energy, while in the line-driven winds studied by \citet{2002ApJ...576..413U}, the kinetic energy is more important than the thermal energy of the wind.

We also show that when the self-consistent interaction of the field lines and the wind is taken into account, the picture of approximately radial streamers (near the star), as observed in the Sun, is not anymore verified for the WTTSs with rotation periods up to $ 10$~d: both open and closed lines are twisted.

With typical parameters of WTTSs, we estimate the time-scale for planet migration due to the torque exerted by the azimuthal ram pressure of the stellar winds, and we showed that for the multi-component corona of a WTTS simulated here, this migration mechanism is not important, in contrast to the results from \citet{2008MNRAS.389.1233L}, who argued that magnetic winds computed using the WD model are expected to be important for planet migration. Further simulations with a greater parameter range (higher magnetic field intensity or higher density at the coronal base) or different system configurations (a tilted stellar magnetosphere or a tilted planetary orbit with respect to the stellar spin axis) may reduce the time-scale to significantly affect the orbital radius of the planet. For instance, if we consider a magnetic field intensity of $2$ - $3$~kG, still consistent with values derived from Zeeman-broadening measurements \citep{2007ApJ...664..975J}, a reduction in $\tau_w$ is expected, as the ram pressure of the wind increases.

\acknowledgments
AAV thanks the Brazilian agencies FAPESP (04-13846-6) and CAPES (BEX4686/06-3). AAV also thanks the warm reception at George Mason University where part of the work was performed. MO acknowledges the support by National Science Foundation CAREER Grant ATM-0747654. MO is also thankful for the hospitality of University of S\~ao Paulo. VJ-P thanks CNPq (305905/2007-4). The authors also thank the anonymous referee for valuable suggestions. The simulations presented here were performed at the Columbia supercomputer, at NASA Ames Research Center.


\end{document}